\documentclass[twocolumn, showpacs, amsmath, amsfontsm, pra]{revtex4}%, galley, pra
\usepackage{graphicx}
\pdfoutput=1
\usepackage{stmaryrd} %for llbracket
\usepackage{aeguill}
\usepackage{fmtcount} % displaying latex counters
\usepackage{dcolumn}% Align table columns on decimal point
\usepackage{bm}% bold math
\usepackage{braket}
\usepackage{pxfonts} %e-print
\usepackage{color}
\usepackage{datetime}
\usepackage[normal]{subfigure} 
\DeclareMathOperator{\sech}{sech}
\usepackage[colorlinks=true, citecolor=blue]{hyperref}
\begin{document}
\title{Ultra-short strong excitation of two-level systems}
\author{Pankaj K. Jha$^{1}$, Hichem Eleuch$^{2}$, Fabio Grazioso$^{3}$\footnote{Email: grazioso@iro.umontreal.ca}}
\affiliation{$^{1}$Department of Mechanical Engineering, University of California, Berkeley, 6141 Etcheverry Hall, CA 94720-1740, USA\\
$^{2}$Department of Physics, McGill University, Montreal (QC), H3A 2TB, Canada\\
$^{3}$DIRO, Universit\'e de Montr\'eal, H3T 1J4, Montreal (QC), H3C 3J7, Canada}
\begin{abstract}
\noindent We present a model describing the use of ultra-short strong pulses to control the population of the excited level of a two-level quantum system. In particular, we study an off-resonance excitation with a few cycles pulse which presents a \emph{smooth phase jump}  i.e. a change of the pulse's phase which is not step-like, but happens over a finite time interval. A numerical solution is given for the time-dependent probability amplitude of the excited level. The  control of the excited level's population is obtained acting on   the shape of the phase transient, and  other parameters of the excitation pulse. 
\noindent 
\end{abstract}
\maketitle
\date{\today}
\section{Introduction}
Ultra-strong pulses with intensities of the order of $10^{14}$ W/cm$^{2}$, and duration of the order of attoseconds, with just few optical cycles, are feasible with present day technology (see e.g. \cite{H1,H2,Goulielmakis-04,Corkum-07,Tsubouchi-08}).
This technological development has been motivated by the large number of possible applications, several of which rely on coherent population transfer techniques. A partial list of such applications is: stimulated Raman adiabatic passage (STI-RAP) \cite{Garcia-05,Zhang-11,Dridi-09,Nakamura-13},  adiabatic rapid passage (ARP) \cite{Jiang-13}, Raman chirped adiabatic passage (RCAP) \cite{Chang-01, Eleuch-12a}, temporal coherent control (TCC) \cite{Brabec-00,Li-13}, coherent population trapping \cite{Harris-97,Issler-10},  optical control of chemical reactions \cite{Yang-10,Cerullo-12},  electromagnetically induced transparency (EIT) \cite{Harris-97,el1,el2,el3,Abdumalikov-10},  efficient generation of XUV radiation \cite{H4,TLA2,TLA3},  breakdown of dipole blockade obtained driving atoms by phase-jump pulses \cite{V3}. Moreover, recently two schemes for efficient and fast coherent population transfer have been presented \cite{Kumar-12}, which use chirped and non-chirped few-cycles laser pulses. Another recent application  \cite{Xiang-12} presents high-order harmonic generation obtained with laser pulses with a $\pi$-phase jump.
Finally, the field of quantum information processing benefits from these results, since many qubit realizations rely on precise quantum levels manipulation \cite{Sete-14,Lee-04,Campbell-10,Grazioso-13,Kim-11}. 
\begin{figure}[b]
	\begin{center}
	\includegraphics[scale=0.7]{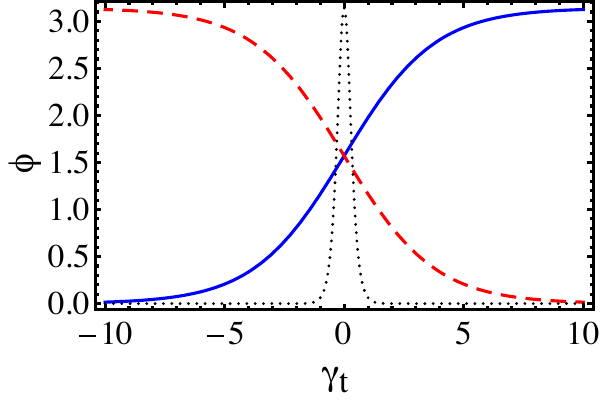}
	\caption{\label{PJ-fig1}The three functional shapes of the smooth phase jumps used: the dashed red is a dropping hyperbolic tangent: $\phi(t) = (\pi/2) [1- \tanh(5 \alpha t )] $, the solid blue is a rising hyperbolic tangent: $\phi(t) = (\pi/2) [1- \tanh(5 \alpha t )] $, and the dotted black one is a hyperbolic secant: $\phi(t) = (\pi/2) \sech(\alpha t ) $. In all the simulations the numerical normalized value is $\alpha=0.265$}
	\end{center}
\end{figure}

The presence of few optical cycles in the pulse gives a constant phase difference between the carrier wave and the pulse shaped envelope \cite{H3}, in contrast with many cycle pulses \cite{V1,V2,V3}. Moreover, optimizing the pulses parameters is proven to enhance the excited state population \cite{R1} or optimizing coherence in two-level systems (TLSs) \cite{R2}. In previous works we have already presented an analytical solution for the dynamics of a TLS excited with pulses of arbitrary shape and polarization \cite{H5,H51}. But since in the model we present here the change rate of levels' populations within a single optical cycle is not negligible, the rotating-wave approximation can't be used. In other words in the present model we can't neglect the contribution of the counter-rotating terms in the Hamiltonian \cite{H5,H51}.
The two levels considered in the  model can be the Zeeman-sublevels and the ultra-short (few to multi-cycle) pulse would be in the radio-frequency regime which has been reported in \cite{h11, JhaCEP1, JhaCEP2}.
% (delete me) is it ok to have the previous line moved here?

We have presented a similar model in another previous work \cite{PrevPRA}, representing the interaction of a TLS with few-cycle pulses, where at time $t=t_{0}$ the phase of the carrier wave jumps of an amount $\phi$, this jump being sharp and step-like. In that work, it has been shown a strong enhancement in the population transfer, for some range of frequencies, with the optimal phase jump of $\phi=\pi$ and the optimal time coincident with the peak of the envelope. 

In the present work we improve that model, considering a \emph{smooth phase change}, i.e. not step-like but happening over a finite interval of time. This new model more closely describes a realistic experimental scenario.

The pulse is characterized by: Rabi frequency $\Omega_{0}$, pulse width $\tau$, carrier frequency $\nu$, phase jump amplitude $\phi$ and phase jump duration $\Delta t$. Moreover, we consider two \emph{qualitative parameters}: the phase jump shape, and the pulse envelope shape. 
We present an analytical solution for the time evolution of the excited state's population, together with a numerical simulation. In the numerical simulation we  use 3 functional shapes for the smooth phase jump: rising hyperbolic tangent, dropping hyperbolic tangent, and gaussian peek, (see Figure \ref{matrix-fig-a}, \ref{matrix-fig-d}, \ref{matrix-fig-g}), whereas for the envelope a gaussian peak has been used. Numerically optimizing the pulses parameters we have obtained enhancements for the population transfer of the order of $10^{4}$.

\section{NUMERICAL SIMULATION}

Be $|a\rangle$ and $|b\rangle$  the states of a two-level atom (TLA), with energy difference $\hbar\omega$, and atomic dipole moment $\wp$. If we let this system interact with a classic field $E(t)={\cal E}(t)\mbox{cos}\nu t$, the equations of motion for the relative wavefunctions are \cite{BK1}:
\begin{subequations}\label{TLA1}
\begin{align}
\dot{C}_{a}&= i\frac{\wp{\cal E}(t)}{\hbar}\mbox{cos}(\nu t) e^{i\omega t}C_{b},\label{second}\\
\dot{C}_{b}&= i\frac{\wp^{*}{\cal E}(t)}{\hbar}\mbox{cos}(\nu t)e^{-i\omega t}C_{a},
\end{align}
\end{subequations}

\begin{figure}[b]
\centering
		{\includegraphics[scale = 0.41]{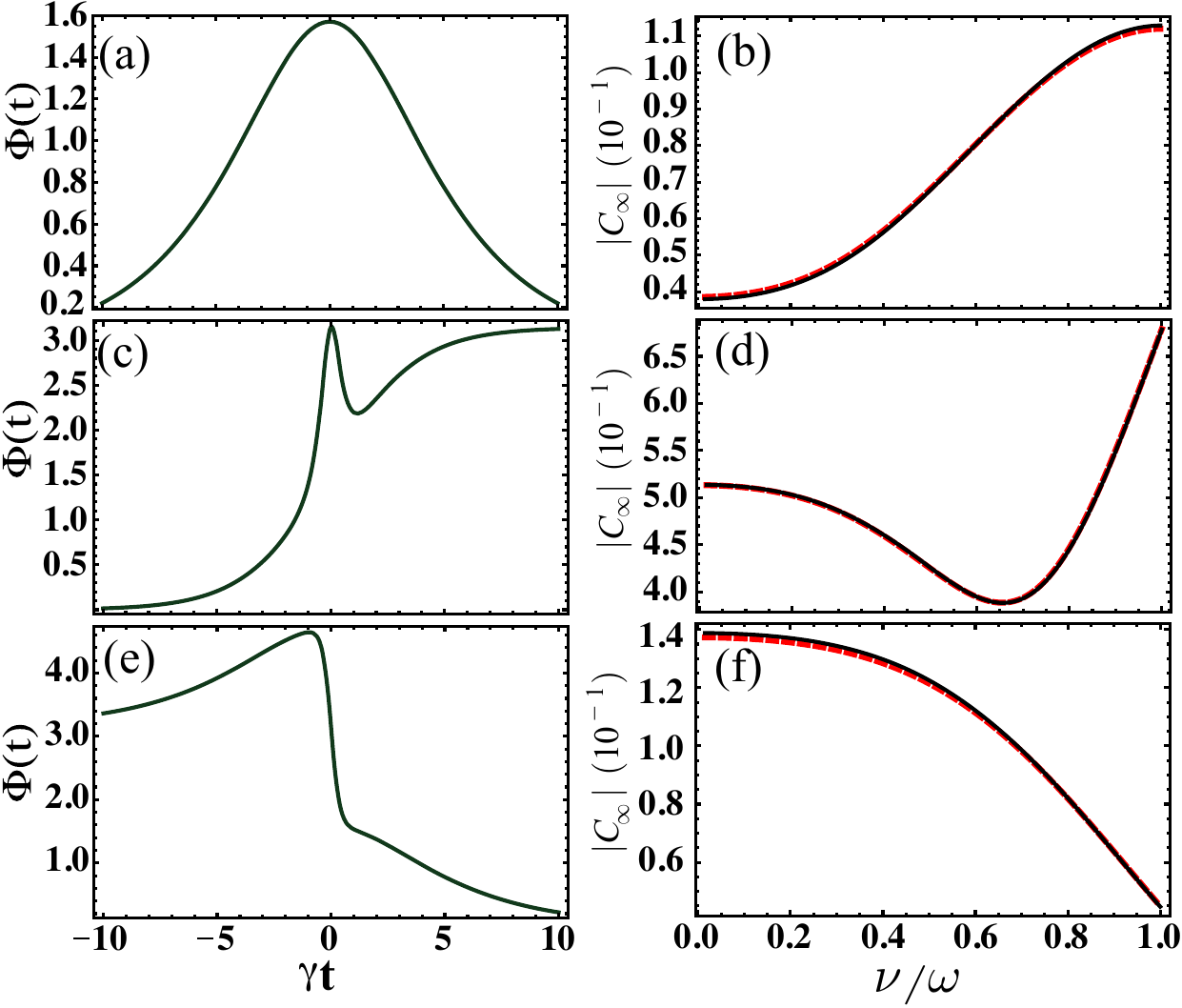}}					
	\caption{\label{comparison-fig}Population left on the upper-level $|a\rangle$ (b,d,f) as a function of ratio of the carrier-frequency ($\nu$) of the excitation pulse and atomic transition frequency ($\omega_{c}$), in the long time limit $t \gg \tau$, for corresponding phase jump function (a,c,e). Here the dashed red line is the numerical simulation of eq.(1) and the solid line is the approximate solution given by eq. \eqref{Riccati-sol}. For the excitation pulse we have used the form $\Omega_{0}(t) = A e^{-\alpha^{2} t^{2}} e^{i \phi(t)}$, and the phase functions have the following forms: (a) $\phi(t)=(\pi/2)\sech[\alpha t]$, (c) $\phi(t)=\pi/2(\sech[10\alpha t]+(1+\tanh[\alpha t])$, (e) $\phi(t)=\pi/2(\sech[\alpha t]+(1-\tanh[10\alpha t]))$. For numerical simulations we chose $A=0.035\omega$, $\alpha=0.265\gamma$ and $\gamma=1.25\omega$ where $\omega=(2\pi)$ 80 Ghz}
\end{figure}

\begin{figure}[!htbp]
	\subfigure[\label{first-label}]
		{\includegraphics[scale = 0.33]{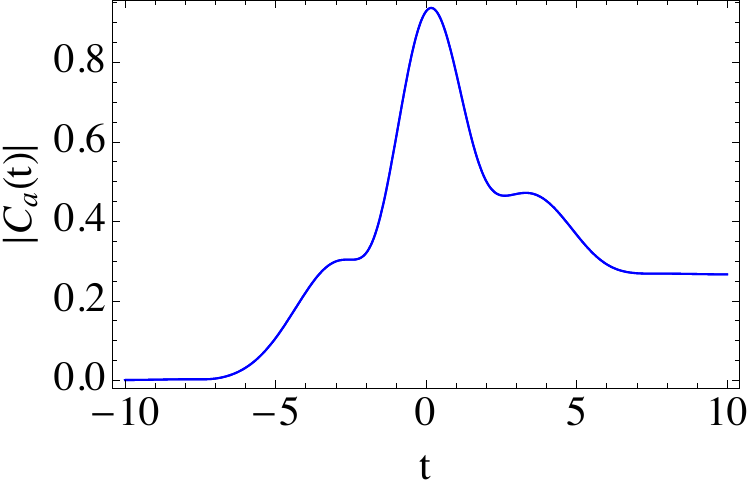}}
		%\hspace{5mm}
	\subfigure[\label{second-label}]
		{\includegraphics[scale = 0.33]{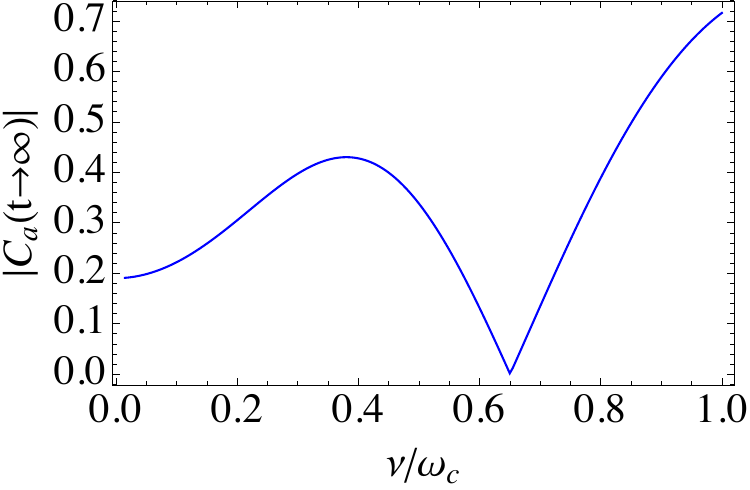}}
	\caption{\label{const-phase-plots} Excited state population in the case of excitation pulse with constant phase. The value used for $\alpha$ is  $\alpha= 0.265$. For numerical simulations we chose $A=0.04375\omega$, $\alpha=0.265\gamma$ and $\gamma=1.25\omega$ where $\omega=(2\pi)$ 80 Ghz. (a) Excited state's population as function of time. - (b) Excited state's population left after the pulse is gone, as a function of the normalized excitation frequency.}
\end{figure}

\begin{figure*}[!htbp]
\centering
	\subfigure[\label{matrix-fig-a}] 
		{\includegraphics[scale = 0.45]{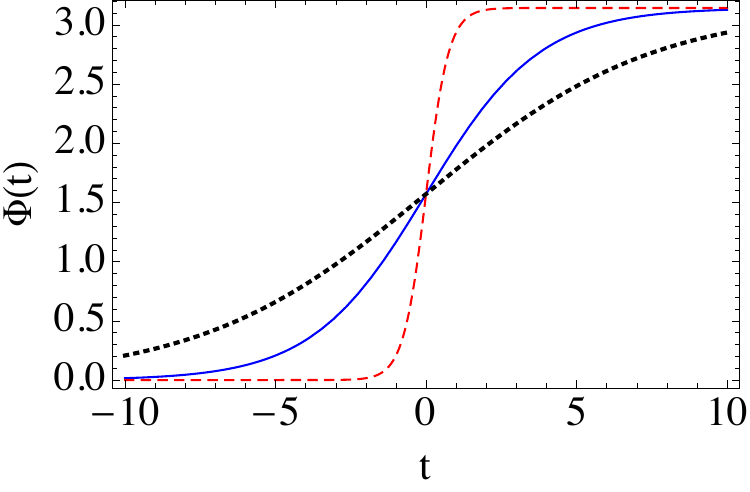}}
	\subfigure[\label{matrix-fig-b} ]
		{\includegraphics[scale = 0.45]{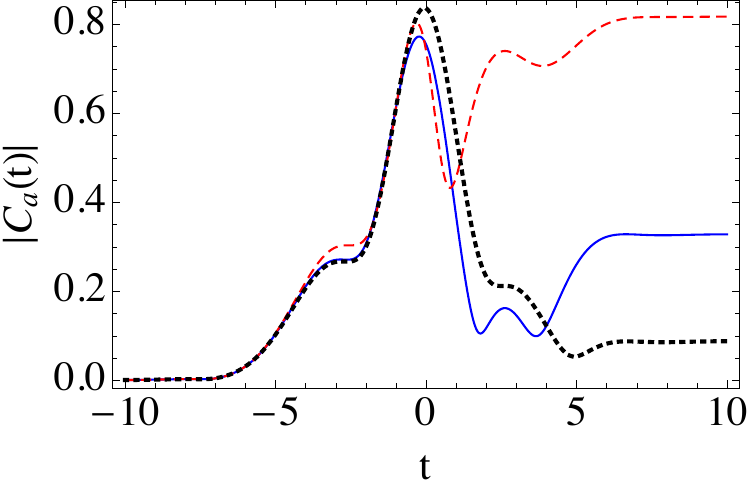}}
		\subfigure[\label{matrix-fig-c} ]
		{\includegraphics[scale = 0.45]{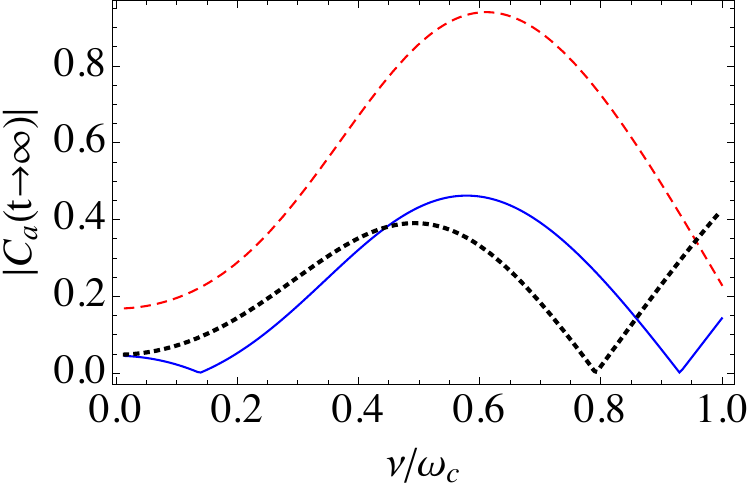}}\\
	\subfigure[\label{matrix-fig-d}]
		{\includegraphics[scale = 0.45]{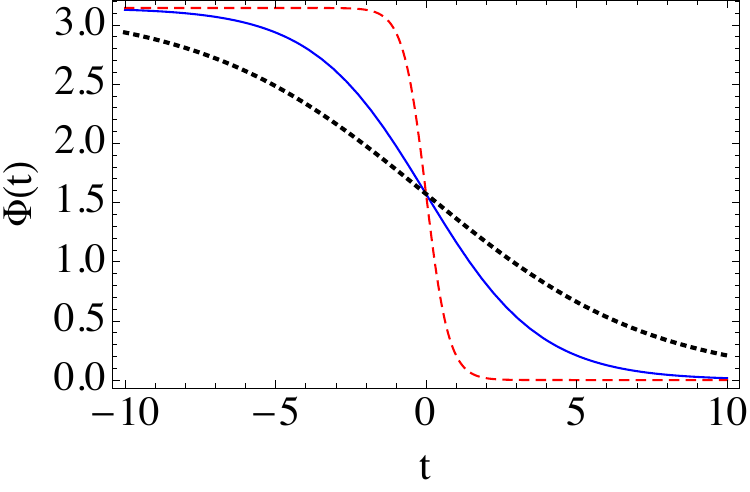}}
	\subfigure[\label{matrix-fig-e} ]
		{\includegraphics[scale = 0.45]{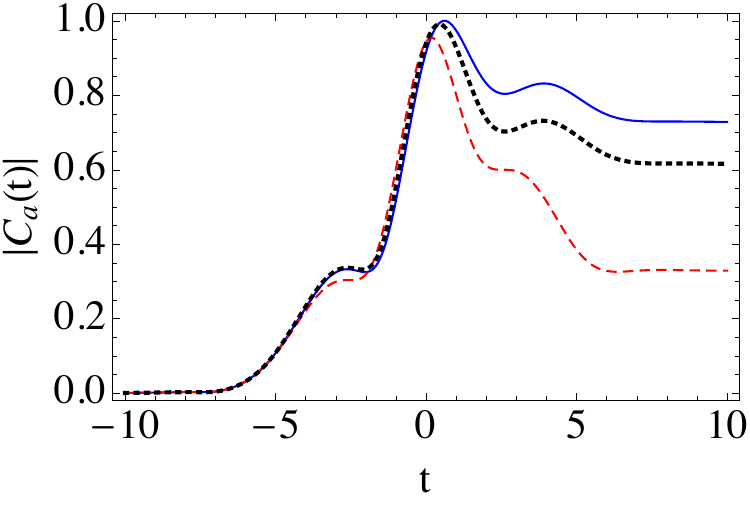}}
			\subfigure[\label{matrix-fig-f} ]
		{\includegraphics[scale = 0.45]{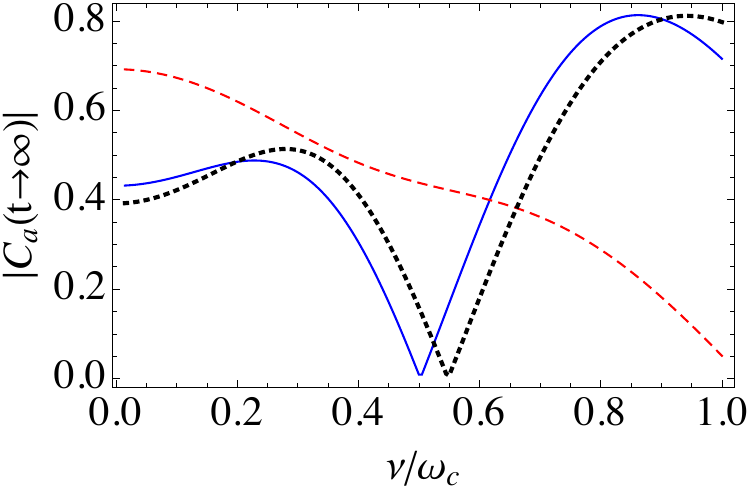}}\\
	\subfigure[\label{matrix-fig-g} ]
		{\includegraphics[scale = 0.45]{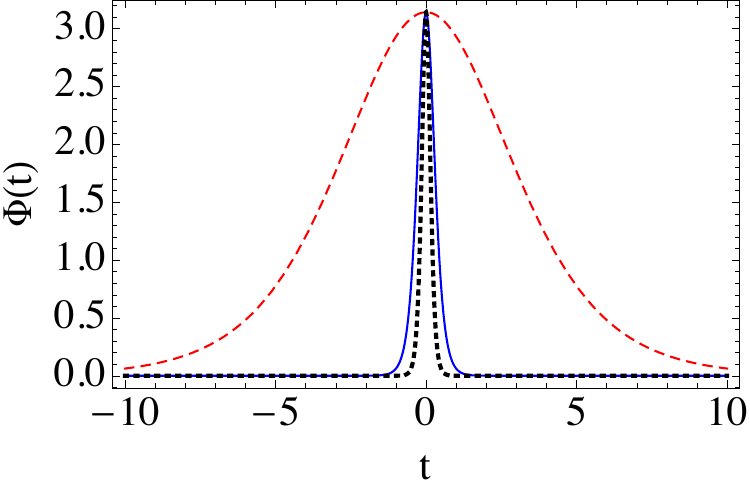}}
	\subfigure[\label{matrix-fig-h}]
		{\includegraphics[scale = 0.45]{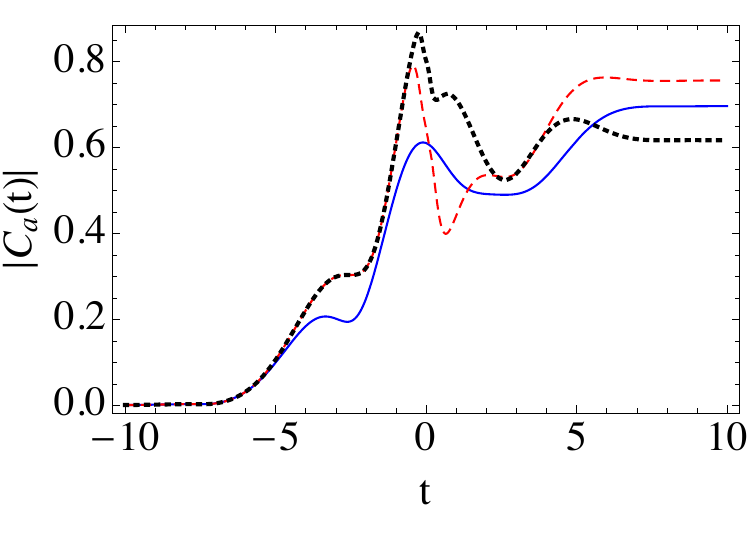}}
			\subfigure[\label{matrix-fig-i}]
		{\includegraphics[scale = 0.45]{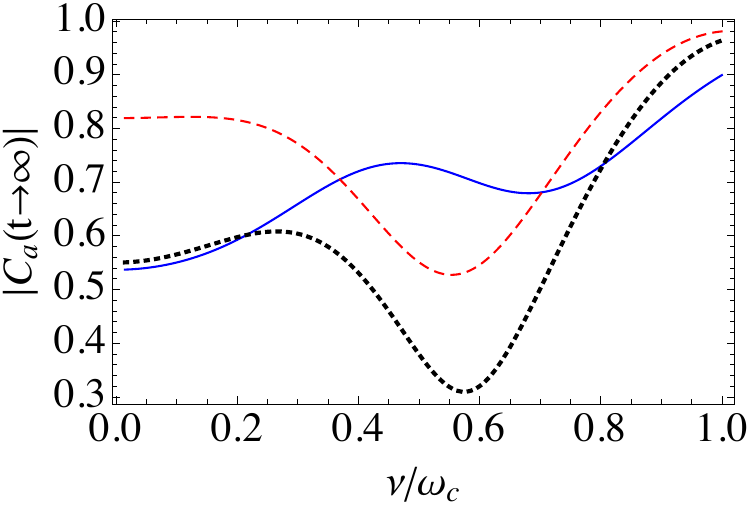}}						
	\caption{\label{simulations-fig} In this figure we present the results of the numerical analysis. Similarly to  \autoref{comparison-fig}, the functional shape of the excitation pulse considered is $\Omega_{0}(t) = A e^{-\alpha^{2} t^{2}} e^{i \phi(t)}$. Each of the three rows of plots refers to a different functional shape of the smooth phase jump (phase-change function $\phi(t)$). For each row we have a plot of the smooth phase jump, a plot of the excited state's population as function of time, and a plot of the excited state's population left after the pulse is gone, as a function of the normalized excitation frequency.  For the plots of the excited state's population as function of time we have used the numerical value of $\nu/ \omega = 0.75$. \\ 
(a)  Phase change of the form   $\phi(t) = (\pi/2) [1+\tanh(\alpha_{1} t)]$, with three different values of $\alpha_{1}$: dashed red (steeper) $\alpha_{1} = 5\alpha$; solid blue (in-between) $\alpha_{1} = \alpha$; dotted black (smoother) $\alpha_{1} = 0.5 \alpha$. -
(b)   Corresponding behaviour of the excited level population $|C_{a}(t)|$. -
(c)    Asymptotic value of the excited state population, as a function of the ``resonance ratio'' (excitation's frequency divided by transition's frequency) for this form of the phase change. -
(d)  Phase change of the form $\phi(t) = (\pi/2)[1- \tanh(\alpha_{1} t)]$, with (as in (a)) three different values of $\alpha_{1}$: dashed red (steeper) $\alpha_{1} = 5\alpha$; solid blue (in-between) $\alpha_{1} = \alpha$; black (smoother) $\alpha_{1} = 0.5 \alpha$. -
(e)  Corresponding behaviour of the excited level population $|C_{a}(t)|$. -
(f)  Asymptotic excited population for this form of the phase change. -
(g)  Phase change of the form: $\phi(t) = (\pi/2) \sech^{2}(\alpha_{1} t)$, with the following values for $\alpha_{1}$: dashed red (larger) $\alpha_{1} = \alpha$; solid blue (in-between) $\alpha_{1} = 10 \alpha$; dotted black (narrower) $\alpha_{1} = 20 \alpha$. -
(h)  Corresponding behaviour of the excited level population $|C_{a}(t)|$. -
(i)  Asymptotic excited population for this form of the phase change. The numerical values used in these plots are the same  as in   \autoref{const-phase-plots}.}
\end{figure*}

\noindent where  $\Delta=\omega-\nu$ is the detuning from resonance. Similarly to \cite{H5, H51}, defining $f(t)=C_{a}(t)/C_{b}(t)$ and $\Omega(t)=\wp{\cal E}(t)/\hbar$, we have the following Riccati equation:
\begin{equation}\label{Riccati-eq}
\dot{f}+i\Omega^{*}(t)\text{cos}(\nu t)e^{-i\omega t}f^{2}-i\Omega(t)\text{cos}(\nu t)e^{i\omega t}=0.
\end{equation}
The approximate solution for eq. \eqref{Riccati-eq}, in terms of the tip angle $\theta$ is given as in \cite{H5}
\begin{equation}
\begin{split}
f(t)&=i\int_{-\infty}^{t}dt'\left\{\left[\frac{d\theta(t')}{dt'}-\theta^{2}(t')\frac{d\theta^{*}(t')}{dt'}\right]\right.\\
&\left. \times \exp\left[2\int_{t'}^{t}\theta(t'')\dot{\theta}^{*}(t'')dt''\right] \right\},
\end{split}
\end{equation}
where the tip angle $\theta(t)$ has been defined as 
\begin{equation}\label{tip-angle}
\theta(t)=\int_{-\infty}^{t}\Omega(t')\text{cos}(\nu t')e^{i\omega t'}dt'
\end{equation}
from which we have $|C_{a}(t)|=|f(t)|/\sqrt{1+|f(t)|^{2}}$. What is of interest is the asymptotic behaviour of $|C_{a}(\infty)|$. In \cite{PrevPRA} is shown good agreement between the analytical and a numerical simulation. To introduce the phase jump, we can write the Rabi frequency as
\begin{equation} \label{modulation-eq}
\Omega(t) = \Omega_{0}(t) \cos \nu t \ e^{i \omega t} e^{i \phi(t)}
\end{equation}
and then, using the same method as in \cite{H5,H51,PrevPRA}, we can obtain an approximated analytic solution for  the Riccati equation \eqref{Riccati-eq}:
\begin{equation} \label{Riccati-sol}
f(t) =i\int_{-\infty }^{t}dt^{\prime }\Phi(t') \text{exp}\left[ 2\int_{t^{\prime }}^{t}\zeta(t'')dt^{\prime\prime }\right],
\end{equation}
The approximate analytical solution is in good agreement with the numerical simulation obtained by directly solving the coupled differential eq. \eqref{TLA1}. From  \autoref{comparison-fig} we see that even for complex phase function the agreement is good. For the sake of completeness, we have added an appendix in which we show the strength of this approach beyond standard TLS. Indeed the Riccati equation approach gives a closed compact form from which both the temporal and steady-state behaviour of the two and three-level system can obtained. 

An interesting observation is that it is possible to rewrite the Rabi frequency in eq. \eqref{modulation-eq} as
\begin{equation}
\Omega(t) = \Omega_{0}(t) \cos \nu t e^{i [\omega t + \phi(t)]}
\end{equation}
and then define $\tilde{\omega}(t) = \omega + \phi(t)/t $ and interpret this as a modulation of the atomic frequency, instead of a modulation of the excitation. Experimentally this can be realized in several ways, e.g. using modulated Zeeman or Stark effect.

Now we move to discuss our numerical simulation of the dynamics of the two-level atom interacting with ultra-short, off-resonant and gradually changing phase $\phi(t)$.

We have computed numerical solutions of the Riccati equation, using different types of phase change (smooth phase jump) functions. In  \autoref{const-phase-plots} we show the case with constant phase, as a reference, and in \autoref{simulations-fig} we show the results of this numerical analysis. The goal of this study is to find the best phase change which allows for the best coupling (most efficient energy exchange) of the excitation pulse with the excited state.

In  \autoref{simulations-fig} each of the three rows of plots refers to a different functional shape of the smooth phase jump (phase change function). For each row we have a plot of the smooth phase jump, a plot of the excited state's population as function of time, and a plot of the excited state's population left after the application of the pulse, as a function of the normalized excitation frequency. Similarly to  \autoref{comparison-fig}, the functional form of the excitation pulse is $\Omega_{0}(t) = A e^{-\alpha^{2} t^{2}} e^{i \phi(t)}$. Moreover, for the plots of  the excited state's population as function of time we have used the numerical value of $\nu/\omega = 0.75$.

\section{Analysis}

For the numerical simulation we have considered the following three phase functions (see  \autoref{simulations-fig}) (a)(b)(c): $\phi(t) = (\pi/2)[1+\tanh(\alpha t)]$, (d)(e)(f): $\phi(t) = (\pi/2)[1-\tanh(\alpha t)]$ and (g)(h)(i) $\phi(t) = (\pi/2)\sech^{2}(\alpha t)$. We can observe a global behaviour which relates the characterizing parameters of the phase change with the amplitude of the population of the excited state. We can see how the phase change duration $\Delta t$, i.e. the steepness of the $\phi(t)$ function, has not an unique effect on the excited population, which depends on the general shape of the phase change. In particular, it is worth noting that for ascending and descending phase changes built on the $\tanh(t)$ function, the effect of the steepness is opposite. Qualitatively, for the ascendent hyperbolic tangent we observe that by increasing the slope the population increases. On the other hand, for the descendent hyperbolic tangent the effect of this parameter is reversed:  decreasing the slope of phase change leads to a decrease of the population. We remark that these behaviours are only global, and are reversed for some small ranges of frequencies. As an example, in plot \ref{matrix-fig-f}, for low ranges of laser frequencies, by decreasing the slope we increase the population, which is opposite of the behaviour observed for higher frequencies. For the peaked shape ($\sech(t)$), no general behaviours are observed. However, for the intermediate range of frequencies it can be observed a link between the increasing of the pulse width and the increase of the population.\\

To mention some quantitative results, with an ascending  $\tanh(t)$, steep phase change (see plot \ref{matrix-fig-c}, dashed red curve, $\alpha_{1} = 5 \alpha$) we achieve an enhancement of $1.9 \times 10^{4}$ in the population transfer, for relative frequency  $\nu/\omega_{c} \sim 0.65$. On the other hand,  near resonant excitation, both $\tanh(t)$ steep phase change generate a remarkable suppression of the excited state population, of a factor of $0.2$ and of $6.8 \times 10^{-2}$ respectively for ascending  $\tanh(t)$ (see plot \ref{matrix-fig-c}, solid blue curve, $\alpha_{1} = \alpha$) and for descending  $\tanh(t)$ (see plot \ref{matrix-fig-f}, dashed red curve, $\alpha_{1} = 5 \alpha$). The bell-shaped $\sech^{2}(t)$ phase change achieves less remarkable results both in enhancement and in suppression,  with an enhancement of a factor $1.4 \times 10^{4}$ for relative frequency of $\nu/\omega_{c} \sim 0.65$ (see plot \ref{matrix-fig-i}, solid blue curve, $\alpha_{1} = \alpha$). In all instances, the value used for $\alpha$ is  $\alpha= 0.265$.

\section{conclusion} %here we report our analytical and numerical results to show the effect of smooth phase jump on the dynamics (both transient and steady-state). 
We have observed that the temporal profile of the phase jump function $\phi(t)$ has a profound effect on the excited state population $|a\rangle$.  We not only can enhance excitation but for the same phase function and other choice of the parameter $\alpha$, we can also suppress it. Such control over excited state dynamics using smooth phase jump as an external parameter can be useful in microwave controlled Raman \cite{Jha13, JhaAPL12}, EIT with superstructures \cite{JhaAPB} to name a few. The approximate analytical solution are in excellent agreement for both delta function \cite{PrevPRA} or smooth phase jump considered here. In appendix \ref{TLA} we present an extension of this approach beyond two-level atom, to three-level in lambda configuration. 
\section{Acknowledgements}
We acknowledge fruitful discussions with Yuri Rostovtsev. P. K. J. acknowledges Herman F. Heep and Minnie Belle Heep Texas A\&M University Endowed Fund held and administered by the Texas A\&M
Foundation and Robert A. Welch Foundation for financial support.

\appendix
\section{Analytical solution for three-level atom} \label{TLA}
\begin{figure}[!t]
	\begin{center}
	\includegraphics[scale=0.4]{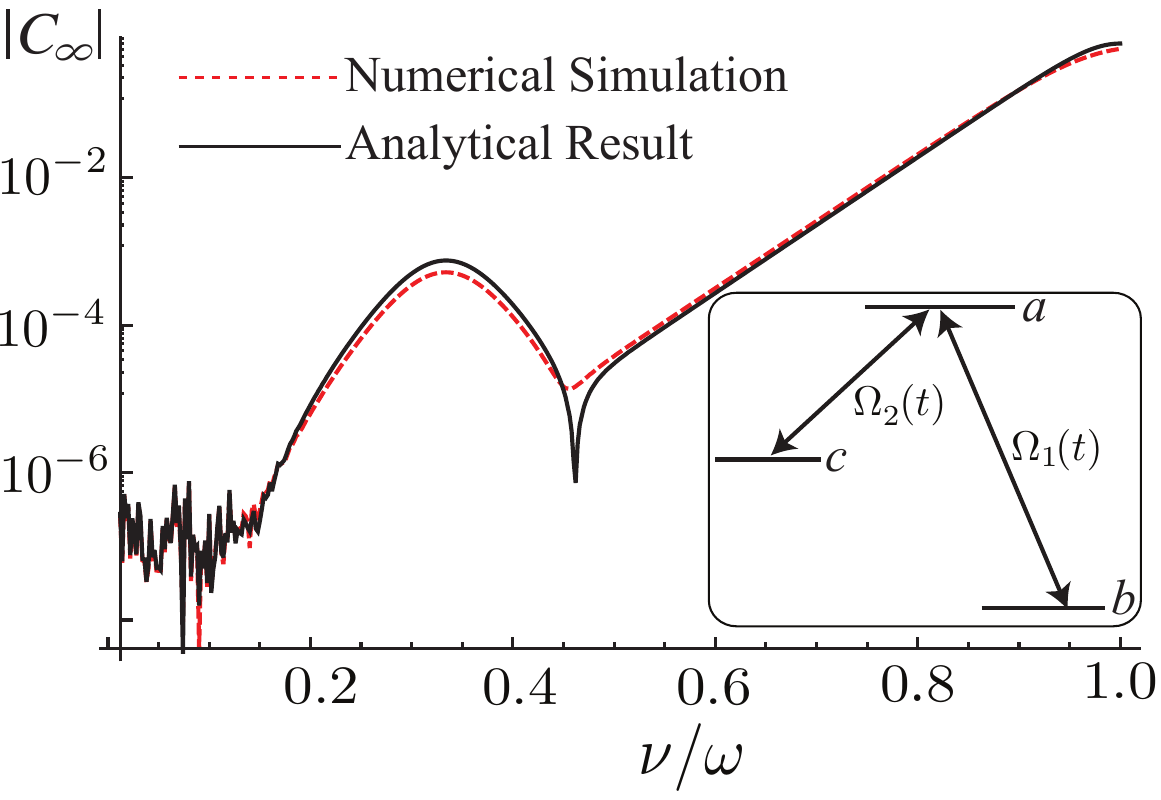}
	\caption{\label{ThLA-fig} Numerical (red dashed line) and analytical (black solid line) solutions of the amplitude of the state $|a\rangle$ after long time
in function of $\nu/\omega$ for the laser pulse envelopes  $\Omega_1 (t)=\Omega_2 (t)=\Omega_0 sech(\alpha t)$. For numerical simulation we chose 
$\Omega_0=.04\omega, \alpha=0.075\omega, \omega_{ab}=\omega_{ac}=\omega=1$.}
	\end{center}
\end{figure}

Motivation to add an appendix on the approximate analytical soltuion for three-level system in lambda configuration is to enlighten the strength of the method used to find the solution for two level atom with and without phase jumps. For the sake of simplicity we will consider constant phase $\phi$. Let us consider a three-level atom(ThLA) in $\Lambda $ configuration [see
 \autoref{ThLA-fig} inset]. The transition $a\leftrightarrow c$ is driven by the field $\Omega
_{2}$, while the field $\Omega _{1}$ couples the $a\leftrightarrow b$
transition. For the time scale considered in this problem, we have neglected
any decays (radiative and non-radiative). The equation of motion for the
probability amplitudes for the states $|a\rangle $, $|b\rangle $ and $%
|c\rangle $ of the ThLA can be written as
\begin{equation}
\dot{C}_{a}(t)=i\tilde{\Omega} _{1}(t)C_{b}(t)+i\tilde{\Omega} _{2}(t)C_{c}(t)  \label{E1}
\end{equation}%
\begin{equation}
\dot{C}_{b}(t)=i\tilde{\Omega} _{1}^{\ast }(t)C_{a}(t)
\label{E2}
\end{equation}%
\begin{equation}
\dot{C}_{c}(t)=i\tilde{\Omega} _{2}^{\ast }(t)C_{a}(t)
\label{E3}
\end{equation}
where $\tilde{\Omega} _{j}(t)$ is defined as the effective
Rabi frequencies
\begin{equation}
\tilde{\Omega}_{j}(t)=\Omega _{j}(t)\cos (\nu
_{j}t)e^{i\omega _{j}t};\qquad j=1,2
\end{equation}
To solve for $C_{a}(t)$ and $C_{c}(t)$ let us define
\begin{equation}
f(t)=\frac{C_{a}(t)}{C_{b}(t)},\,\,\,
g(t)=\frac{C_{c}(t)}{C_{b}(t)}
\end{equation}%
In terms of $f(t)$ and $g(t)$ Eqs.(\ref{E1}, \ref{E2}, \ref{E3}) reduces to
\begin{equation}
\dot{f}(t)+i\tilde{\Omega}_{1}^{\ast }f^{2}(t)=\tilde{\Omega} _{1}+i\tilde{\Omega}\Omega _{2}g(t)
\label{E4}
\end{equation}
\begin{equation}
\dot{g}(t)+i\tilde{\Omega}_{1}^{\ast }f(t)g(t)=i\tilde{\Omega}_{2}^{\ast}f(t)  \label{E5}
\end{equation}
In order to solve these equations we extended the method developed \cite{H5,H51}. By neglecting the non-linear term $f^{2}(t)$ and the term $\propto g(t)$ in
eq.(\ref{E4}) we can solve for $f_{1}(t)$ as
\begin{equation}
f_{1}(t)=i\int_{-\infty }^{t}\tilde{\Omega}_{1}dt^{\prime }
\end{equation}%
Similarly by neglecting the term $\propto g(t)$ in eq.(\ref{E5}) we can
solve for $g_{1}(t)$ as
\begin{equation}
g_{1}(t)=-\int_{-\infty }^{t}\tilde{\Omega}_{2}^{\ast}%
(t^{\prime })\theta _{1}(t^{\prime })dt^{\prime }  \label{E6}
\end{equation}%
where the tip angle $\theta _{1}(t)$ is defined as
\begin{equation}
\theta _{1}(t^{\prime })=\int_{-\infty }^{t}\tilde{\Omega}_{1}(t^{\prime })dt^{\prime }
\end{equation}%
Next let us write the non-linear term in eq.(\ref{E4}) as
\begin{equation}
f^{2}(t)=\left[ f(t)-f_{1}(t)\right] ^{2}+2f(t)f_{1}(t)-f_{1}^{2}(t)
\end{equation}%
Then eq.(\ref{E4}) can be written as
\begin{equation}
\begin{split}
\dot{f}(t)+i\tilde{\Omega}_{1}^{\ast
}(t)\{[f(t)-f_{1}(t)]^{2}+2f(t)f_{1}(t)-f_{1}^{2}(t)\}\\
=i\tilde{\Omega}_{1}(t)+i\tilde{\Omega}_{2}(t)g(t)  \label{E7}
\end{split}
\end{equation}
Let us assume that $g(t)\approx g_{1}(t)$ and we neglect $[f(t)-f_{1}(t)]^{2}$ \cite{H5} in this case we can write eq.(\ref{E7}) in term of the tip angles $\theta _{1}(t)$ and $\theta _{2}(t)$
\begin{equation}
\dot{f}(t)+i\dot{\theta}_{1}^{\ast }(t)\left\{
2f(t)f_{1}(t)-f_{1}^{2}(t)\right\} =i\dot{\theta}_{1}(t)+i\dot{\theta}%
_{2}(t)g_{1}(t)  \label{E8}
\end{equation}%
where
\begin{equation}
\theta _{2}(t^{\prime })=\int_{-\infty }^{t}\tilde{\Omega}_{2}(t^{\prime })dt^{\prime },
\end{equation}%
The analytical solution of the equation eq. (\ref{E8}) is then:
\begin{equation}
f(t)=e^{-a(t)}\int_{t_{0}}^{t}b(t^{\prime })e^{a(t^{\prime })}dt^{\prime }
\end{equation}
where
\begin{equation}
a(x)=2i\dot{\theta}_{1}(t)f_{1}(t)
\end{equation}
and
\begin{equation}
b(x)=i\dot{\theta}_{1}(t)+i\dot{\theta}_{2}(t)g_{1}(t)+i\dot{\theta}%
_{1}^{\ast }(t)f_{1}^{2}(t)
\end{equation}
For $g(t)$ the solution can be obtain from eq.(\ref{E5})where we use $f(t)\approx f_{1}(t):$
\begin{equation}
\dot{g}(t)+i\dot{\theta}_{1}^{\ast }(t)f_{1}(t)g(t)=i\dot{%
\theta}_{2}^{\ast }(t)f_{1}(t)
\end{equation}
which give us 
\begin{equation}
g(t)=e^{-c(t)}\int_{t_{0}}^{t}D(t^{\prime })e^{c(t^{\prime })}dt^{\prime }
\end{equation}
where
\begin{equation}
c(x)=i\dot{\theta}_{1}^{\ast }(t)f_{1}(t)
\end{equation}
and
\begin{equation}
D(x)=i\dot{\theta}_{2}^{\ast }(t)f_{1}(t)
\end{equation}
 In  \autoref{ThLA-fig} we have plotted the numerical (red dotted line) and analytical (blue solid line) solutions of the amplitude of the state
$|a\rangle$ after long time in function of $\nu/\omega_c$ for the laser pulse envelopes  $\Omega_1 (t)=\Omega_2 (t)=\Omega_0
\sech(\alpha t)$ with $\Omega_0=.04\omega, \alpha=0.075\omega, \omega_{ab}=\omega_{ac}=\omega=1$. We see that the approximate analytical solution matches well with the numerics under the parameters considered here. Extension of this methodology to Schrodinger equation \cite{H6,H7} and position dependent mass Schrodinger(PDMSE) equation can be found in \cite{JhaJMO11,Hichem3D}.


\begin{thebibliography}{99}
\bibitem{H1} M. Wegener, Extreme Nonlinear Optics: An Introduction
(Springer, Berlin, 2005).
\bibitem{H2} T. Brabec and F. Krausz, Rev. Mod. Phys. 72, 545 (2000).
\bibitem{Goulielmakis-04}
E. Goulielmakis, M. Uiberacker, R. Kienberger, A. Baltuska,
V. Yakovlev, A. Scrinzi, T. Westerwalbesloh,
U. Kleineberg, U. Heinzmann, M. Drescher, et al., Science
305, 1267 (2004).
\bibitem{Corkum-07}P. B. Corkum and F. Krausz, Nat Phys 3, 381 (2007).
\bibitem{Tsubouchi-08} M. Tsubouchi, A. Khramov, and T. Momose, Phys. Rev. A 77, 023405 (2008).
\bibitem{Garcia-05} R. Garcia-Fernandez, A. Ekers, L. P. Yatsenko, N. V. Vitanov, and K. Bergmann, Phys. Rev. Lett. 95, 043001 (2005).
\bibitem{Zhang-11} B. Zhang, J.-H. Wu, X.-Z. Yan, L. Wang, X.-J. Zhang, and J.-Y. Gao, Opt. Express 19, 12000 (2011).
\bibitem{Dridi-09} G. Dridi, S. GuÈrin, V. Hakobyan, H.R Jauslin and H. Eleuch,  Phys. Rev A 80, 043408 (2009).
\bibitem{Nakamura-13}S. Nakamura, H. Goto, and K. Ichimura, Optics Communications 293, 160 (2013), ISSN 0030-4018.
\bibitem{Jiang-13}L.-J. Jiang, X.-Z. Zhang, G.-R. Jia, Z. Yong-Hui, and X. Li-Hua, Chinese Physics B 22, 023101 (2012).
\bibitem{Chang-01}B. Y. Chang, I. R. Sol√°, V. S. Malinovsky, and J. Santamar√≠a,Phys. Rev. A 64, 033420 (2001).

\bibitem{Eleuch-12a} H. Eleuch, S. Guerin, and H. R. Jauslin, Phys. Rev. A 85, 013830 (2012).

\bibitem{Brabec-00}T. Brabec and F. Krausz, Rev. Mod. Phys. 72, 545 (2000).
\bibitem{Li-13} Y. Li, Y. Zhang, C. Li, and X. Zhan, Optics Communications 287, 150 (2013).
\bibitem{Harris-97} S. E. Harris, Physics Today 50, 36 (1997).
\bibitem{Issler-10} M. Issler, E. M. Kessler, G. Giedke, S. Yelin, I. Cirac, M. D. Lukin, and A. Imamoglu, Phys. Rev. Lett. 105, 267202 (2010).
\bibitem{Yang-10} X. Yang, Z. Zhang, X. Yan, and C. Li, Phys. Rev. A 81, 035801 (2010).
\bibitem{Cerullo-12} G. Cerullo and C. Vozzi, Physics 5, 138 (2012).
\bibitem{el1}  H. Eleuch, and R. Bennaceur, Journal of Optics A: Pure and Applied Optics 5, 528 (2003). 
\bibitem{el2} N. Boutabba, H. Eleuch and H. Bouchriha, Synthetic Metals 159, 1239 (2009).
\bibitem{el3} H. Eleuch, D. Elser, and R. Bennaceur, Laser Phys. Lett. 1, 391 (2004). 
\bibitem {Abdumalikov-10} A. A. Abdumalikov, O. Astafiev, A. M. Zagoskin, Y. A. Pashkin, Y. Nakamura, and J. S. Tsai, Phys. Rev. Lett. 104, 193601 (2010).
\bibitem{H4} E. A. Sete, A. A.  Svidzinsky, Y. V.  Rostovtsev, H. Eleuch, P. K. Jha, S. Suckewer, and M. O. Scully, IEEE J. Sel. Top. Quantum Electron. 18, 541 (2012).
\bibitem{TLA2} P. K. Jha and Y. V. Rostovtsev, Phys. Rev. A 81, 033827 (2010)
\bibitem{TLA3} P. K. Jha and Y. V. Rostovtsev, Phys. Rev. A 82, 015801 (2010).    
\bibitem{V3} J. Qian, Y. Qian, M. Ke, X.-L. Feng, C. H. Oh, and Y.Wang, Phys. Rev. A 80, 053413 (2009).
\bibitem{Kumar-12} P. Kumar and A. K. Sarma, Phys. Rev. A 85, 043417 (2012).
\bibitem{Xiang-12} Y. Xiang, Y. Niu, H. Feng, Y. Qi, and S. Gong, Opt. Express 20, 19289 (2012).
\bibitem{Sete-14} E. A. Sete, H. Eleuch,  Phys. Rev. A 89, 013841 (2014).
\bibitem{Lee-04} K. F. Lee, D. M. Villeneuve, P. B. Corkum, and E. A. Shapiro, Phys. Rev. Lett. 93, 233601 (2004).
\bibitem{Campbell-10} W. C. Campbell, J. Mizrahi, Q. Quraishi, C. Senko, D. Hayes, D. Hucul, D. N. Matsukevich, P. Maunz, and C. Monroe, Phys. Rev. Lett. 105, 090502 (2010).

\bibitem{Grazioso-13} F. Grazioso, B.R. Patton, P. Delaney, M.L. Markham, D.J. Twitchen, and J.M. Smith, Appl. Phys. Lett. 103, 101905 (2013).

\bibitem{Kim-11} D. Kim, S. G. Carter, A. Greilich, A. S. Bracker, and D.Gammon, Nat. Phys. 7, 223 (2011).



\bibitem{H3} A. Baltuska, T. Udem, M. Uiberacker, M. Hentschel, E. Goulielmakis, C. Gohle, R. Holzwarth, V. Yakovlev, A. Scrinzi, T. Hansch, et al., Nature 421, 611 (2003).
\bibitem{V1} N. V. Vitanov, N. J. Phys. 9, 58 (2007).
\bibitem{V2} B. T. Torosov and N. V. Vitanov, Phys. Rev. A 76, 053404 (2007).
\bibitem{R1} N. Dudovich, D. Oron, and Y. Silberberg, Phys. Rev. Lett. 88, 123004 (2002).
\bibitem{R2} S. Malinovskaya, Optics Comm. 282, 3527 (2009).
\bibitem{H5} Y. V. Rostovtsev, H. Eleuch, A. Svidzinsky, H. Li, V. Sautenkov, and M. Scully, Phys. Rev. A 79, 063833 (2009).
\bibitem{H51}  Y. V. Rostovtsev, and H. Eleuch, J. Mod. Opt. 57, 1882 (2010). 
\bibitem{h11}H. Li, V. A. Sautenkov, Y. V. Rostovtsev, M. M. Kash, P. M. Anisimov, G. R.Welch, and M. Scully, Phys. Rev. Lett. 104, 103001 (2010).
\bibitem{JhaCEP1} P. K. Jha, H. Li, V. A. Sautenkov, Y. V. Rostovtsev, and M. O. Scully, Opt. Commun. \textbf{284}, 2538 (2011).
\bibitem{JhaCEP2}P. K. Jha, Y. V. Rostovtsev, H. Li, V. A. Sautenkov, and M. O. Scully, Phys. Rev. A \textbf{83}, 033404 (2011).
\bibitem{PrevPRA} P. K. Jha, H. Eleuch, and Y. V. Rostovtsev, Phys. Rev. A 82, 045805 (2010).
\bibitem{BK1} M. O. Scully and M. S. Zubairy, Quantum Optics (Cambridge University Press, Cambridge, England, 1997).
\bibitem{Jha13}P. K. Jha, S. Das and T. N. Dey (unpublished) arXiv:1210.2356
\bibitem{JhaAPL12}P. K. Jha, K. E. Dorfman, Z. Yi, L. Yuan, Y. V. Rostovtsev, V. A. Sautenkov, G. R. Welch, A. M. Zheltikov, and M. O. Scully, Appl. Phys. Lett. \textbf{101},  091107 (2012).
\bibitem{JhaAPB}P. K. Jha and C. H. R. Ooi (unpublished) arXiv:1205.5262
\bibitem{H6} H. Eleuch, Y. V. Rostovtsev, M. O. Scully, EPL 89, 50004 (2010).

\bibitem{H7} H. Eleuch and Y. V. Rostovtsev, J. Mod. Opt. 57, 1877 (2010).
\bibitem{JhaJMO11}P. K. Jha, H. Eleuch, and Y. V. Rostovtsev, J. Mod. Opt. 58, 652 (2011).
\bibitem{Hichem3D}H. Eleuch, P. K. Jha, and Y. V. Rostovtsev, Math. Sci. Lett. 1, 1 (2012).
\end{thebibliography}
\end{document}